# Image-based Communication on Social Coding Platforms

**Maleknaz Nayebi**[1*]    |    **Bram Adams**[2*]

[1]EXINES Lab, York University, Canada

[2]MCIS Lab, Queen's University, Canada

**Correspondence**
Email: mnayebi@yorku.ca

**Funding information**

Visual content in the form of images and videos has taken over general-purpose social networks in a variety of ways, streamlining and enriching online communications. We are interested to understand if and to what extent the use of images is popular and helpful in social coding platforms. We mined nine years of data from two popular software developers' platforms: the Mozilla issue tracking system, i.e., Bugzilla, and the most well-known platform for developers' Q/A, i.e., Stack Overflow. We further triangulated and extended our mining results by performing a survey with 168 software developers. We observed that, between 2013 and 2022, the number of posts containing image data on Bugzilla and Stack Overflow doubled. Furthermore, we found that sharing images makes other developers engage more and faster with the content. In the majority of cases in which an image is included in a developer's post, the information in that image is complementary to the text provided. Finally, our results showed that when an image is shared, understanding the content without the information in the image is unlikely for 86.9% of the cases. Based on these observations, we discuss the importance of considering visual content when analyzing developers and designing automation tools.

Coding, Data Science, Software Analytics

# 1 | INTRODUCTION

Visual content, as shared on image-based social networks such as Snapchat and Instagram, has transformed day-to-day online communication since it allows communication in a visual way for events and messages that would otherwise be tedious to describe textually (for example, people's lunch, feelings, or travel experiences). According to Hootsuite news[1], in July 2022, Instagram reported 54 million users and overtook Facebook to claim second place in the worldwide rankings of social networks. The number of Snapchat users was four times higher in 2018 in comparison to 2014, and in 2018 over 200 million snaps were being shared every day. During the same period, on Instagram the number of users had grown five-fold, with more than 95 million images shared on Instagram.

The use of visual content for interacting on social media is not exclusive to the general public. For over a decade, social coding platforms that host daily professional activities of developers have been studied. It has been discussed that the social behavior of developers largely follows the trend of societal behavior [1], and their behaviors on these platforms have been the subject of numerous studies. Diagrams borrowed from the UML specification [2], architectural standards [3], or software processes [4] are widespread, and so are the omnipresent whiteboard scribblings used during brainstorming discussions of modern software engineering teams [5]. However, despite the growth in sharing visual content by the general public, traditionally, our tools and techniques for assisting software developers or supporting their decisions have evolved around textual content [6, 7, 8, 9]. In fact, the existing social coding platforms and the automated software engineering tools are unable to retrieve, process, and extract information from developers' shared images. For example, bugs or user questions involving GUI elements cannot be queried by developers or linked to specific code review discussions or code locations. This is unfortunate since the use of visual content by developers in the context of their daily tasks provides an opportunity to make better decisions by broadening the information. We believe that the increasing trend in the amount of visual content shared by developers might eventually make automated decision-making in software engineering impossible without mining the corresponding image data [10].

In this study, we perform three exploratory analyses to understand the state of the practice in terms of visual content in software-engineering communication. First, we analyze the state of image sharing by the developers of Mozilla by analyzing their Bugzilla issue tracking system, which is one of the most popular bug repositories in the software engineering research community [11], over a period of nine years. Then, we mine Stack Overflow as a popular Q/A platform for developers by taking a close look at the images shared on these platforms during the same period of nine years. Third, we performed a qualitative analysis of the nature of the images shared on these platforms and performed a qualitative study with the developers to understand their perception of the importance and helpfulness of image sharing.

This paper is an extension of an ICSE 2020 NIER paper [10]. Compared to the ICSE 2020 NIER paper [10], this paper provides further evidence on the status and trends of sharing images across two popular software development media. We (i) added four years of data into analysis throughout the study compared to the original paper, (ii) separated the analysis of trends in image sharing for questions from those for responses on Stack Overflow, (iii) we now present results of two additional survey questions with 168 software developers (in **RQ2**), (iv) new to this version, we performed further analysis on the relation between images and other attributes of posts on Q/A platforms and issue tracking systems. Our results support the vision of visual social coding environments, intending to encourage the community to extend tools and techniques. In particular, we address the below research questions:

**RQ1:** What is the trend of sharing images in Bugzilla and Stack Overflow?

We looked into the trend of image sharing between 2013 to 2022. We gathered and analyzed all the bug reports across Mozilla's core, Firefox Android, and Firefox iOS projects. Also, we retrieved images from Q/A discussions with hashtags of Python, Java, or JavaScript programming languages.

**RQ2:** What is the content (nature) of images shared by software developers?

Three experienced developers categorized a sample of 2,000 images to identify potential categories. 165 other developers then categorize 22,561 images from Bugzilla and Stack Overflow into the ten identified categories.

**RQ3** How do developers perceive the value of the shared images in software teams?

We performed a survey with 168 developers about their behavior toward visual content on social coding and Q/A platforms. We further analyzed the value of sharing images by asking them to indicate whether images provide additional information to the text, and if it would be possible to understand a bug report or the Q/A without the image. We overall evaluated the value of 22,561 images to achieve this goal.

**RQ4:** How does the presence of images correlate with the communities' responsiveness?

We used descriptive statistics in addition to lightweight qualitative observations to explore the relationship between different attributes of issue reports and Q/A posts, such as the length of response, number of replies, time to get the first reply, and time to get the resolution.

**RQ5:** How does the content (nature) of images relate to other image attributes and posted issues?

We performed further in-depth analysis by evaluating the relationship between the type of image (as defined in **RQ2**) and the responsiveness attributes discussed in **RQ4**. We intended to evaluate if a particular content or image type tends to be helpful for the developers compared to the others.

This paper is exploratory and observatory in its nature. We performed statistical tests on the quantitative data from Bugzilla and Stack Overflow. We used these to design a survey study with software developers to understand their perception of the value of the shared images qualitatively. The analysis of the ratio between the textual and visual content of Bugzilla's change requests and in Q/As of programming questions on Stack Overflow revealed a steady increase in sharing images over the past eight years. Developers' shared images are meaningful and provide complementary information compared to their associated text. The shared images are often essential in understanding the change requests, questions, or responses submitted. It is crucial for the reader to interpret these results within their limitations and not establish a causal relationship.

In what follows, we first (Section 2) provide an overview on the related work. Then in Section 3, we discuss our methodology for answering each research question. We follow by explaining our dataset and the process of gathering information about Bugzilla and Stack Overflow as well as the developers' survey in Section 4. In Section 5.1, we discuss **RQ1** and the trend of image sharing between 2013 to 2022. In Section 5.3, we answer **RQ2** by surveying developers, and in Section 5.2 we report the results of image categorization and image types identified by the survey developers in Response to **RQ3**. We report our results on **RQ4** in Section 5.4 by analyzing confounding factors. In Section 5.5, we deepen our analysis for each issue type to answer **RQ5**. We wrap up the paper by discussing the limitations and threats to the validity (Section 6), and the conclusion (Section 7). To start off, in the next section (Section **??**) we provide a brief overview of the state-of-the-art with regards to image sharing and processing in software engineering and the areas that can be extended and changed by this emerging concept.

## 2 | RELATED WORK

With the emergence of modern version control systems such as GIT and the tools and techniques for improving the productivity of software developers, the discussion on multiple aspects of social coding platforms has been of particular interest to the research community. The earlier version of this study [10] has been the first to investigate the impact of image sharing on these platforms. We briefly overview the state-of-the-art directions and techniques for social coding in Section 2.1.

In Section 2.2, we provide an overview of the problems the software engineering research community is currently addressing using image processing techniques. To the best of our knowledge, at the time of writing this study, none have focused on social coding platforms.

### 2.1 | Developer productivity and social coding

In 2005, Zimmermann et al. [12] suggested the process of mining version histories to guide software changes. Bagel et al. [1] elaborated on the impact of social media on software teams and development processes and the emergence of Q/A platforms. Later, Dabbish et al. [13] offered the term social coding in 2012 for the large-scale, transparent, and distributed collaboration of software developers facilitated through version control systems such as GitHub. Following this, Gousios et al. [14] analyzed pull-based software development. All these initiated and further strengthened the focus of the software engineering community on the analysis of developers' contributions and communications, both individually and in teams. These research efforts led to a deeper analysis of commit messages, release notes, pull request messages, bug reports, etc., which all use natural language processing [15, 16] to analyze the textual communication in the software teams. To the best of our knowledge, these studies use quantitative contribution indicators such as the extent of code change, number of releases, etc., primarily focusing on the textual communications from and between the developers [17, 18, 19].

There are a number of systematic and comprehensive literature studies that tried to summarize these efforts. Huo et al. [20] elaborated further on the use of deep learning in bug localization, and Polisetty et al. [21] discussed the essence and usefulness of these advanced machine learning techniques for localizing bugs. A more comprehensive overview provided by Zakari et al. [22] compares all techniques used for localizing faults. In none of these studies the use of non-textual communications has been pointed out or discussed.

The use and analysis of Stack Overflow for software development also have an extensive history. Yet, none of the existing works focuses on non-textual communications or shared media, which is the main emphasis of this study. Following the study of Begel et al. [1], Squire et al. [23] analyzed the utility of social media for the software development process. These early works have been mostly focused on social indicators such as likes, upvotes, etc., as the early stages of social media evolved around "connection", while these days, the differentiating advantage of these media relates to the different forms of "communication" [24, 25], which is motivating this study. Ahmad et al. [26] provided an overview of the efforts for better usage of social Q/A platforms such as Stack Overflow. Recently, these authors extended their study [27] to provide an overview of the deep learning techniques for mining and analyzing Stack Overflow. Yet, despite the advances in deep learning techniques for image processing, all these efforts are limited to analyzing textual information.

## 2.2 | Image processing in software engineering

The current use of image analysis in software engineering is mainly focused on software testing to detect duplicate test reports in crowd-testing platforms, leading to an increase in the number of such publications in recent years.

Studies showed a high number of duplications in crowd-testing reports submitted to crowd-testing platforms (Wang et al. [28] reported 82% average duplication). When performing crowd testing, the test reports tend to be a screenshot with a short description. Liu et al. [29] studied these test reports for mobile apps by retrieving and refining the description of similar screenshots. In a preliminary study, they tested their proposed approach on four industrial, mobile, and crowdsourced projects and demonstrated acceptable performance for the practitioners. To reduce triaging effort and to increase the precision of duplicated test report detection, Wang et al. [28] combined the image structure feature and image color feature with textual descriptions of crowd test reports. Their suggested method, which combines screenshots and textual descriptions, outperformed the text-based methods for detecting duplicate reports. The CroReGcan method proposed by Yu et al. [30] generates bug reports in crowd-sourced mobile testing platforms. The method analyzes bug scenarios in the form of app screenshots using widget analysis and performs screenshot transitions. The generated crowd-sourced testing bug reports for mobile apps showed successful preliminary results.

Zhao et al. [31] elaborated on the linting of GUI animations against design-don't guidelines and approached the problem as a multi-class screencast classification task which, due to the lack of labeled data, has been approached through unsupervised learning. In order to improve GUI testing, White et al. [32] proposed a method for identifying GUI widgets in screenshots using machine learning. Using this data resulted in an average increase of 42.5% in performance compared to conventional random testing. Coppola et al. [33] provided a comprehensive overview and comparison between GUI testing techniques, including the image-based methods. Chen et al. [34] proposed a new deep learning-based model for detecting GUI elements and applied it to mobile applications. In the same direction, Chen et al. [35] introduced LabelDroid to predict the labels of image-based buttons automatically. Cooper et al. [36] used image processing to detect duplicate video-based bug reports. They used visual and textual input to evaluate 180 bug reports across six Android apps to detect duplicate bug reports. However, this work is mainly focused on video screencasts of apps' GUI compared to the former studies, which focused on processing images only.

Following the habits of software developers in inspecting test interactions with an application through its GUI, Stocco et al. [37] suggested a method called VISTA for visual test repair. They suggested a fast image processing pipeline to validate test cases visually. They tested VISTA on 2,672 tests spanning four subject systems to collect a dataset of 733 individual test breakages. VISTA could resolve 81% of such breakages. In another study, Stocco et al. [37] showed that on average, 2.8% of Layout-based and 3.9% of Visual test methods are fragile and in need of maintenance. Around 20% of Layout-based and 30% of Visual test, suites had to be modified at least once within an app's life cycle. Choudhary et al. [38] introduced WebDiff to identify cross-browser issues using a visual comparison of the appearance of a rendered web page, along with a structural analysis of the page's DOM. This led to a series of studies for learning image features [39] and image-based web testing [40, 41].

Screencasts are recordings of a user or expert showing how to use one or more features of a given application. Since such screencasts show not only the major user-visible features but also their workflow through a sequence of button clicks, user input, etc., they have been found useful for deriving a wide range of information about the underlying code base. For example, Ponzanelli et al. [42, 43] use screencasts to identify video fragments that are most relevant to developers using a specific API. Moslehi et al. [44] establish traceability links between use case scenarios demonstrated in a screencast and the source code implementing those scenarios.

# 3 | EMPIRICAL METHODOLOGY

This paper is exploratory in nature. Exploratory research involves studying objects within their natural context and allowing the findings to emerge from observations [45]. This necessitates the use of a flexible research design that can adapt to changes in the observed phenomenon. In our study, we employed a mixed method strategy, utilizing both qualitative and quantitative data. It combines qualitative research and data to interpret a phenomenon based on explanations provided by individuals while complementing it with quantitative observations for better generalization. After observing the existing trends in **RQ1**, we investigated the developers' perception and understanding of the problem being investigated in **RQ2** and **RQ3**. To gain insights into individuals' perceptions, we conducted a survey as an established empirical method [45, 46]. This helped us understand and report the context of image sharing in software teams. Further, we explored the relationship with the variables in **RQ4** and **RQ5**.

In this section, we provide an overview of the empirical protocols utilized for addressing **RQ1** to **RQ5**. For **RQ1**, we collected data from Stack Overflow and Bugzilla and conducted a descriptive analysis to examine the trends in image sharing. **RQ2** was addressed through surveys and interviews with users, focusing on the content of the shared images. **RQ3** involved gathering additional data, including user demographics and contextual information, to explore the perceived value of image sharing on social coding platforms. **RQ4** and **RQ5** were involved further investigation of the relationship between image sharing, utilizing statistical methods to analyze the data. These protocols ensured a systematic approach to investigating each research question, resulting in evidence-based findings regarding image sharing on Stack Overflow and Bugzilla.

## 3.1 | Trends across Time (RQ1)

Our primary research question focuses on investigating the trends over time of image sharing on both Stack Overflow and Bugzilla. This question serves as an observatory and descriptive analysis, providing valuable insights into the current practices and facilitating future research endeavors. In order to address this question, we collected data as outlined in the next section and subsequently organized it by timestamp, discretizing the information into yearly intervals. Utilizing this organized dataset, we generated informative visualizations and conducted descriptive statistical analyses to present our findings.

## 3.2 | Survey with Software Developers (RQ2 and RQ3)

After observing software developers increasing tendency to share images, we were interested in exploring if these images are informative and essential to understanding a post (either a change request or Q/A). In **RQ2** we fur-ther investigated the nature of the shared images to better understand what type of information is being shared by the developers. Then, in **RQ3**, we performed a survey study with software developers, as we are interested in un-derstanding developers' perceptions. Survey provides a comprehensive system for collecting information to describe attitudes and behavior of software developers in sharing images [46].

To find expert developers answering this task, we first posted a task on the BountySource platform[2] and linked it to an equivalent task in Amazon Mechanical Turk (MTurk)[3] in a way that only developers having an access code posted on BountySource would be able to participate in our study. 168 developers volunteered to participate in our study. These developers had at least two years of professional experience and have been active in open-source communities. In **RQ2**, we hired three software developers among volunteers for participants in our study with the most years of experience. These developers had a minimum of seven years of coding experience to identify categories for the content of images. For each image, each developer answered the following question:

- What does the image communicate?

---

[2] https://salt.bountysource.com/
[3] https://www.mturk.com

At first, developers could openly define and name the categories of the type of information the image is communicating and could classify each image into multiple classes. Then, we asked the developers to peer-review the categories of another developer and revise their own categorization if needed. Finally, one of the paper's authors reviewed all the categories and aggregated them into ten categories with the highest frequency of occurrence. The developers answered this question for 2,000 randomly selected images across Bugzilla and Stack Overflow. With these categories established, we started the second round of crowd-sourcing by asking developers to categorize more images. This dataset did not overlap with the data set of the first round of the analysis. We asked the rest of the 168 developers (165 developers) who volunteered to participate in our study to categorize each image in at least one of the categories defined in the first round or categorize them as "other".

We then performed a more holistic survey with the developers to answer **RQ3**. We first asked the developers four general questions:

- How often do you respond to the issues or posts of other developers on issue tracking systems (such as Bugzilla) or Q/A platforms? (Five-point Likert scale)
- How often do you post content including an image on issue tracking systems (such as Bugzilla) or Q/A platforms? (Five-point Likert scale)
- Do you prefer to respond to issues or questions with or without visual content? (prefer content with an image, neutral, prefer content without an image)
- Why do developers share images along with the text? (answer chosen between precision, time to respond, and comprehension, or "other reasons")

Then, we provided the textual body of the Stack Overflow questions and replies or of the bug reports. For each image, we asked *three* software developers per image to answer the below questions:

- Does the image provide additional information to the text? (Yes/No question)
- How likely would you understand the text without having access to the image? (Five-point Likert scale)

## 3.3 | Relationships between attributes (RQ4 and RQ5)

We conducted an analysis to measure the responsiveness of Bugzilla developers and the activity of developers on Stack Overflow, using various attributes as proxies. However, it is essential to note that these attributes differ between the two platforms. For Bugzilla, we compared reports with and without an image based on several factors: (i) the textual length of the issue, (ii) the number of replies, (iii) the type of issue, (iv) the status of the issue, and (v) the time taken to close the issues. We analyzed these attributes both in general and separately for each issue type, status, and severity. Regarding Stack Overflow, we compared posts that included an image with those that did not, considering attributes such as: (i) the length of the question, (ii) the length and number of answers, (iii) the number of answers received, (iv) the acceptance of answers, and (v) the time taken to receive the first answer.

To compare the posts with and without images, we utilized the Mann-Whitney test. The Mann-Whitney test is a non-parametric statistical test used to investigate differences between the group means of different attributes. In our analysis, we applied the Mann-Whitney test [47] to examine attributes such as $Time_{Closure}$, the number of replies, and other relevant factors previously mentioned. This test is particularly suitable for comparing two independent groups when the data may not follow a normal distribution or when the assumptions of parametric tests cannot be met. By employing the Mann-Whitney test, we were able to determine whether there were significant differences between the two groups in terms of the analyzed attributes. The test provided insights into the potential variations and relationships between the presence of images and the measured outcomes, enabling us to draw meaningful conclusions from our analysis. We also calculated and reported Cliff's delta effect size [48] for these statistical tests.

Moving into **RQ5**, we tested eight hypotheses using Kruskal-Wallis tests to compare the group means of eight factors for the ten types of images defined in **RQ2** (also see Figure 3). We tested the below hypotheses:

$H_{01}$: There is no difference in the group means of *# of replies in Bugzilla* for the different image types.

$H_{02}$: There is no difference in the group means of *# of types of issues in Bugzilla* for the different image types.

$H_{03}$: There is no difference in the group means of *# of status of issues in Bugzilla* for the different image types.

$H_{04}$: There is no difference in the group means of *severity of issues in Bugzilla* for the different image types.

$H_{05}$: There is no difference in the group means of $Time_{Closure}$ *in Bugzilla* for the different image types.

$H_{06}$: There is no difference in the group means of *# of answers in Stack Overflow* for the different image types.

$H_{07}$: There is no difference in the group means of *# of accepted answers* for the different image types.

$H_{08}$: There is no difference in the group means of *the time to get a first answer* for the different image types.

While there is no assumption about the sample size in the ANOVA test, the unequal sample sizes reduce the robustness and the statistical power of this test [47]. As a result, we opted for a Kruskal-Wallis test (also known as the one-way ANOVA test), which is considered a non-parametric version of the ANOVA test. The results of a Kruskal-Wallis test did not show any significant difference for six of the eight hypotheses comparing the group means of the factors listed in $H_{01}$ to $H_{08}$.

## 4 | EMPIRICAL DATA AND PROCESS

To comprehensively understand the use of images among software developers, we performed both quantitative and qualitative analysis on Bugzilla issue tracking data of Mozilla and on Q/A data of Stack Overflow. Below, we explain the data used in our analysis. To form a representative set for our qualitative analysis of the images, we randomly chose 50% of the images shared in Bugzilla between 2013 and 2022. We also randomly chose 10% of the images retrieved from Stack Overflow in this period. This resulted in 5,187 images from Bugzilla and 17,374 images from Stack Overflow. The 22,561 images in total formed the data we used for answering **RQ3**. To understand the content of the shared images, we formed a data set by randomly choosing 2,000 images across Bugzilla and Stack Overflow. The 2,000 images were randomly selected and evenly distributed between Stack Overflow and Bugzilla over time (between 2013 to 2022) and over a programming language (Python, Java, and JavaScript), respectively, for each type of Mozilla project (Core, Firefox, iOS, and Android). We used the data set of **RQ2** containing 22,561 Images (10% of Stack Overflow posts[4] and 50% of Bugzilla change requests).

---

[4]Posts refer to both the question and the subsequent replies.

## 4.1 | Bugzilla Data

Bugzilla has been a major source for the software engineering research community to develop automated tools and techniques for bug triaging, localization, and other innovative automation to assist software developers. For this reason, we used Bugzilla as one of our studied cases in this paper.

In particular, we gathered change requests and replies from Mozilla's Core, Firefox, Firefox for Android, and Firefox for iOS projects[5]. We decided to perform a ten-year trend analysis within the years in which image-based social networks gained increasing popularity[6]. In our preliminary observation, we found that before 2013 the number of images shared on Bugzilla and Stack Overflow was not considerable enough for a meaningful side-by-side comparison of trends in these two platforms. Hence, we used Bugzilla's REST API to gather all the issues and their available attributes over the nine years between January 2013 to January 2022[7] This resulted in 65,611 bug reports across Mozilla projects.

For **RQ1**, we used the text body of issue reports and replies, the time stamp of each issue report and reply, as well as the type and status of the issue reports. Based on Bugzilla's official description[8], the issue reports in Bugzilla have three types:

**Defect:** Including regression, crash, security issue, and any other reported issue,

**Enhancement:** Including new features, UI improvement, performance, and other requests for changes visible to the end user, excluding engineering changes (such as design),

**Task:** Including refactoring, removing, replacing, and enabling of functionality and any other engineering tasks.

According to Bugzilla guidelines, each issue report can have one of the eight statuses Assigned, Closed, Needinfo, New, Reopened, Resolved, Unconfirmed, or Verified. However, in the data we gathered for this study, we only had one of the below four statuses, and the rest did not exist in our sample:

**New:** Issues that just opened and are still considered not well elaborated by developers, with no other status defined for them,

**Reopened:** Issues that have been assigned and fixed by a developer, but someone else has reproduced the problem again,

**Assigned:** Issues that a developer is currently working on; this could be a bug report or an enhancement,

**Unconfirmed:** The issue has been submitted but not confirmed or reproduced by other developers.

Also, each issue report includes information about its severity[9], i.e., the impact of a bug. By default, the severity value is empty and is further defined when triaging the issue. If not applicable, the value will be changed to N/A; otherwise, one of the four values below is assigned to each issue, as quoted from Mozilla's Wiki:

---

[5] https://bugzilla.mozilla.org/describecomponents.cgi

[6] https://www.hootsuite.com/

[7] The original NIER paper was published in 2019 which included a five year period. We updated the dataset to include the last four years.

[8] https://wiki.mozilla.org/BMO/UserGuide/BugFields#bug_type

[9] https://wiki.mozilla.org/BMO/UserGuide/BugFields#bug_severity

**Catastrophic (S1):** Blocks development/testing, may impact more than 25% of users; causes data loss, likely dot release driver, and no workaround available,

**Serious (S2):** Major functionality/product severely impaired, and a satisfactory workaround does not exist,

**Normal (S3):** Blocks non-critical functionality, and a workaround does exist,

**Small/Trivial (S4):** Minor significance, cosmetic issues, low or no impact to users.

## 4.2 | Stack Overflow Data

Stack Overflow is often analyzed as a source of crowd-sourced information for software development. The software engineering research community has been focused on developing tools and techniques to improve the Q/A process and knowledge retrieval and consolidation. Hence, we used the data from this platform to observe the trend and status of crowd-sourced image sharing by developers. To retrieve posts with images, we used the Stackexchange REST API[10]. For the period of January 2013 to January 2022 (including nine years of data), Stack Overflow included 1,391,224 threads associated with one of the tags Python, Java, or JavaScript. These threads included 3,364,177 posts overall.

We then gathered the text, date of posting, date of first reply, images (all if more than one is available in a post), and number of answers. For answering **RQ4** and **RQ5**, we used the textual body of the Q/As in Stack Overflow to investigate the length of questions and the length and number of responses for each. We also separated the accepted answers from the rest of each question and extracted the timestamp of each question and response.

## 5 | RESULTS

This section responds to each of the five research questions in turn.

## 5.1 | Trend of Image Sharing Over Time (RQ1)

**The proportion of Bugzilla reports and Stack Overflow posts sharing images follows a steady increase from 2013 to 2022.** Figure 1 shows the number of change requests in Bugzilla for the four studied Mozilla projects between 2013 to 2022. Looking at the number of posts with images relative to the total number of posts on Bugzilla (34,1143 posts), there has been an increasing trend over the past nine years. The trend shows a significant increase when considering only the main change requests (excluding follow-up replies and posts). In 2018, the number of change requests with images doubled compared to 2013 (a 200% increase), while the total number of change requests only increased by 20%. It's worth noting that Mozilla's iOS application was launched in 2015, and there were no change requests prior to that time.

For Stack Overflow, Figure 2 plots the number of posts with/without images and the ratio of the number of posts including an image out of the total number of posts per year. This analysis has been shown for the overall data and for each programming language. The results show a linearly increasing trend in the number of Stack Overflow posts that include an image. Our analysis also showed a (linearly) increasing overall trend in the proportion of the number of questions, including an image in Stack Overflow. Motivated by these observations, we went a step further to seek developers' own perception of sharing images in their daily tasks.

---

[10] https://archive.org/details/stackexchange

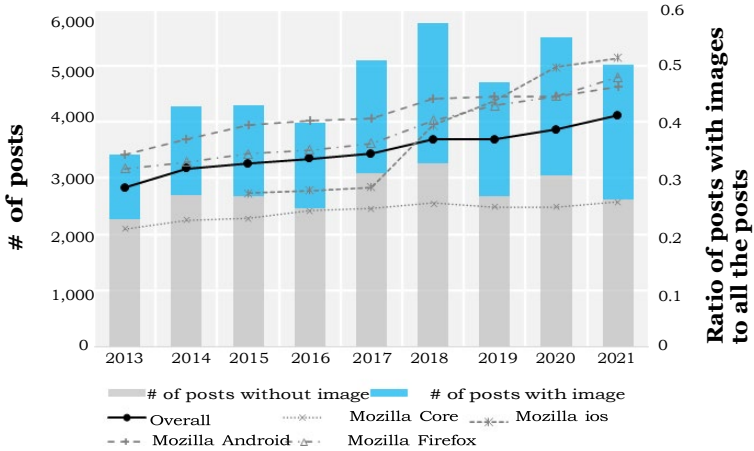

**FIGURE 1** Trend of sharing images in four popular Mozilla projects between 2013 to 2022.

**FINDING 1:** There is an increasing trend in sharing images on social coding platforms - In other words, the growing trend of people sharing visual content also applies to developers in the context of software development processes and tasks.

**So What?** Looking into the trend on image-based social media platforms and relying on the studies in software engineering that report on the social behavior of software teams [1, 49], has led us to believe that this increasing trend will be continuing. While both Stack Overflow and Bugzilla allow users to attach and share images, they are not fundamentally designed for visual communications. This is a core difference between these platforms and the more general-purpose ones like Instagram and Snapchat.

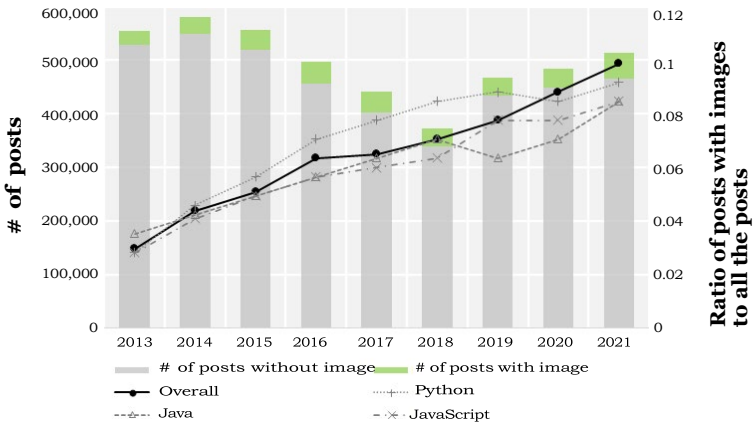

**FIGURE 2** Trend of sharing images in Stack Overflow between 2013 to 2022 for three popular programming languages.

## 5.2 | Content of the Shared Images (RQ2)

As detailed in Section 4, we hired three developers to participate in the survey. We asked them to perform open sorting [50] to categorize a randomly chosen subset of 2,000 images across Stack Overflow and Bugzilla. **As the result of the first round of crowd-sourcing, we ended up with ten categories of images added to posts**, namely (examples are provided in Figure 3):

**Code and IDE:** Screenshot of a code snippet in an IDE or shell command,

**Run time error:** Screenshot of the run time error, either only the message or also the menus and IDEs involved,

**Menus and preferences:** Screenshot of the menus and preferences in the development or deployment environments,

**Program input:** Screenshot of spreadsheets/documents or an illustration to communicate the input of a software system,

**Desired output:** An illustration of the output that software developers desire to achieve,

**Program output:** Screenshot of the output of an executed code snippet,

**Dialog box:** Screenshot of dialog messages such as crash reports, system errors, IDE update messages, etc.,

**Steps and processes:** Screenshot of the system or an illustration created by developers to demonstrate the steps of achieving a particular goal,

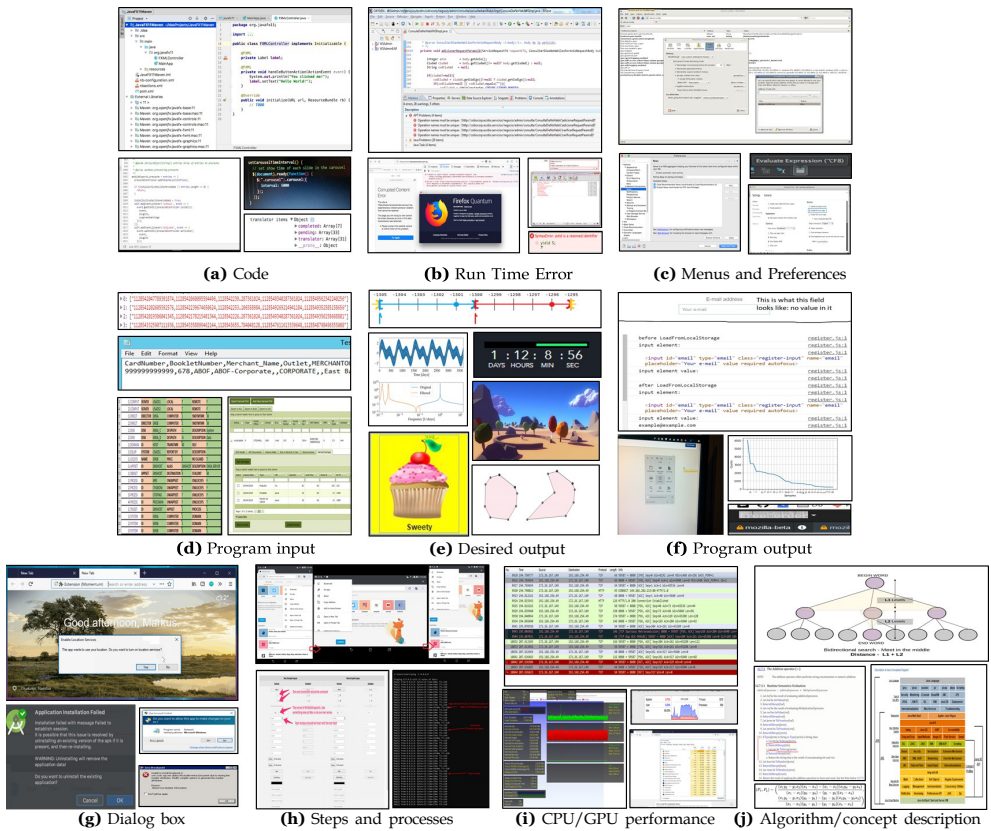

(a) Code    (b) Run Time Error    (c) Menus and Preferences

(d) Program input    (e) Desired output    (f) Program output

(g) Dialog box    (h) Steps and processes    (i) CPU/GPU performance    (j) Algorithm/concept description

**FIGURE 3** Sample images posted by developers in Bugzilla or Stack Overflow (further details in Table 3).

**CPU/GPU performance:** Screenshot of the system performance,
**Algorithm and concepts:** Screenshot of diagrams, tutorials, algorithms, pseudo-code, etc.

During the categorization and unification process followed by the three developers, the level of Cohen's Kappa agreement between the developers was 0.83, demonstrating a *very good* level of agreement [48]. As the result of categorizing 22,561 images in this second round of crowd-sourcing, we gathered the frequency of images of each type using 165 other developers volunteering to participate in our study. Table 1 shows the categories and the frequency of images within each category.

**The majority (62.8%) of the images contained *code and IDE*, followed by 42.5% of the images showing Run time errors.** While the proportion of these images differed between Stack Overflow and Mozilla, the top two categories were the same in both data sources. For Stack Overflow, the following three most common categories are Steps and process (35.1%), Program output (20.8%), and Menus and preferences (16.9%), while for Bugzilla, the next three categories are Dialog box (23.3%), Menus and preferences (21.3%) and Program output (13.7%). We noticed significant discrepancies for the Dialog box (1.7% for Stack Overflow vs. 23.3% for Bugzilla), Steps and processes (35.1% vs. 7.9%), and Algorithm/concept (16.2% vs. 0.1%). We believe that most of these differences are due to the domain of both platforms, i.e., Bugzilla contains posts regarding issues of Mozilla's end-user applications (e.g., Firefox web browser). At the same time, Stack Overflow covers a wide range of posts on API usage, etc.

**TABLE 1** Frequency of each type of image in Stack Overflow, in Bugzilla, and Overall.

| Category | Overall | | Stack Overflow | Bugzilla |
|---|---|---|---|---|
| Code | 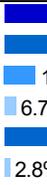 | 62.8% | 69.4% | 40.5% |
| Run time error | 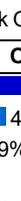 | 42.5% | 46.2% | 30.3% |
| Menus and preferences | 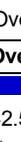 | 17.9% | 16.9% | 21.3% |
| Dialog box | 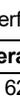 | 6.7% | 1.7% | 23.3% |
| Steps and processes | 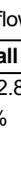 | 28.8% | 35.1% | 7.9% |
| Program input | 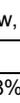 | 2.8% | 3.7% | 0.06% |
| Desired output | 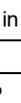 | 9.2% | 10.6% | 4.6% |
| Program output | 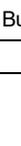 | 19.1% | 20.8% | 13.7% |
| CPU/GPU performance | 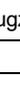 | 2.8% | 3.18% | 1.4% |
| Algorithm/concept | 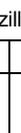 | 12.5% | 16.2% | 0.1% |

## 5.3 | Developers' Perception of the Shared Images (RQ3)

168 developers participated in our crowd-sourced tasks for evaluating 22,561 images. These developers had a minimum of two years of professional experience and have been active in open source communities. Table 2 shows the initial questions to understand the demographics and preferences of the participated developers. We further showed developers a selected sample of images (as discussed in Section 4) and asked them three questions. Table 3 summarizes the results of the analysis based on the majority of developers' votes per image. As the results of the survey with developers (**RQ2**) showed, images still represent information complementary to text and are not standalone content being shared by the developers.

**The surveyed developers are highly familiar with issue tracking systems and Q/A platforms.** The majority of them (87.1%) has sometimes or more often posted on issue tracking systems or Q/A platforms. 72.1% of them sometimes or more often replied to other developers' posts. Among the latter, 58.9% voted positively when asking them if they have a preference toward responding to content that includes an image.

**TABLE 2** Interaction frequency and preference of the 168 surveyed developers.

| How often do developers post on issue tracking systems or Q/A platforms? | |
|---|---|
| Never | 0.6% |
| Rarely | 12.5% |
| Sometimes | 57.2% |
| Always | 22.7% |
| Often | 7.2% |

| How often do developers react to issues or posts of other developers? | |
|---|---|
| Never | 3% |
| Rarely | 25% |
| Sometimes | 31.6% |
| Always | 39.3% |
| Often | 1.2% |

| Do developers prefer to interact with contents with or without an image? | |
|---|---|
| With visual content | 58.9% |
| Neutral | 22.4% |
| Without visual content | 18.7% |

**TABLE 3** Results of surveying the informativeness of 22,561 images using crowdsourcing with 168 developers.

| Does the image provide additional information comparing to text? | |
|---|---|
| Yes | 87.8% |
| No | 12.2% |

| How likely can you understand the text without the image? | |
|---|---|
| Very Likely | 2.7% |
| Likely | 10.4% |
| Unlikely | 46.9% |
| Very Unlikely | 40.0% |

| Why do developers share images along with text? | |
|---|---|
| To be more precise in describing the problem/solution | 70.8% |
| To be faster in describing a problem or responding | 82.7% |
| To facilitate comprehending or responding to the problem | 89.9% |
| Other reasons | 10.1% |

Out of the 22,561 studied images, the majority (87.8%) was identified as informative, and developers considered 86.9% of the corresponding posts unlikely or very unlikely to comprehend without the image. In contrast, developers considered 12.2% of the images (8,244 images) uninformative, and for 13.1% images, they considered the associated text likely or very likely to be understandable without the image. In general, most of the surveyed developers (89.9%) use images to help comprehend the text. 82.7% also used it to be faster in describing and posting. 70.8% of the developers considered images useful for being more precise about the details. 10.1% selected other reasons and described it as *being more fun*, and *being catchy*. For instance, one developer stated:

*"I want my question to stand out amongst the flood of other [questions]. I sometimes add a meme to make it catchy!"*
Another developer mentioned:

*"Images make it easier to understand, and it doesn't hurt to make it a more enjoyable time for others who are helping out*

*answering my, sometimes dumb, questions."*

**FINDING 2:** Images provide information complementary to the text - Images can provide additional information in social coding environments compared to text. Despite the increasing grace of visual content on social media and social coding environments, one would need to demonstrate if and to what extent the information communicated via images is useful compared to the text and the associated metadata. Our survey with the developers showed that 87.8% of them found complementary information in the accompanying images and that for 86.9% of them, the text was not informative enough without the image. However, it should be noted that not all the images are equally informative. For example, screenshots of a code snippet, found in **RQ2** to be the most common type of shared image included in Stack Overflow and Bugzilla posts, could be redundant with the text provided by a developer.

> **So What?** Mining non-textual information on software development platforms for understanding developers' behaviors and automating development-related tasks should be a focus for the research community.
>
> As developers (increasingly) communicate by images, we should be able to automatically retrieve, infer, and decide based on non-textual information. Considering the results of our analysis and change trends in social media towards image-intensive networks [51], we foresee that, gradually, images might become a prominent means of sharing information in developers' social coding environments. A short text followed by an image might become a new, popular format of a bug report, change request, question, or response.
>
> While mining such visual information currently is not performed, it is not unfeasible. Image analysis is the process of extracting meaningful information from images by using automated techniques and has evolved into multiple sub-fields, including but not limited to image recognition, image segmentation, motion detection, image captioning, and color recognition. The advances in convolutional neural networks made object detection, image captioning and summarization, generation of unauthentic images and holograms (moving images), and sentiment and color analysis of images all possible, with reliable accuracy. This field is growing fast, with multiple applications in self-driving cars, health and science, fashion, and homeland security and defense systems. Hence, similar techniques could be considered as a starting point for dealing with image analysis in the context of software engineering data.

## 5.4 | Relation Between Sharing Images and the Community's Responsiveness (RQ4)

We first discuss the question for Bugzilla and follow by the analysis of Stack Overflow.

### 5.4.1 | Responsiveness on Bugzilla

**On average, the length of the issue reports without an image is 46% longer than the issues with an image, yet this difference is not statistically significant.** We started by looking into the length of the text in terms of the number of words or characters in the posts with and without images. Figure 4 shows the distribution of the length of issue reports. The average length of an issue report with an image attachment is 349.6 characters and 48.7 words, compared to an average of 554.3 characters and 100.4 words for posts without any images. This shows that the length of the issue reports without an image is, on average, twice (46% more characters and 60% more words) the size of the issues with an image. However, the median of their length is just slightly different (delta = $0.69$), leading the difference in the distribution of the length both in terms of characters and words for the issue reports with and without images to not be statistically significant (Mann-Whitney test p-value for characters = $0.6$ and for words = $0.3$ for an $\alpha = 0.05$).

When looking into the number of replies, we observed that **the number of responses for the issue reports includ-**

**TABLE 4**  Number of different types of issues with and without image data in Bugzilla for 2013-2022.

|  | Defect | Enhancement | Task |
|---|---|---|---|
| **With image** | 18,251 (91.4%) | 1,701 (8.5%) | 18 (0.1%) |
| **Without image** | 39,619 (81.4%) | 8,302 (17.1%) | 744 (1.5%) |

**ing an image, is on average two times more than the number of issue reports without an image**. When comparing the distributions, the Mann-Whitney test shows a significant difference in the p-value = $0.001$ (for an $\alpha = 0.05$) with Cliff's delta effect size = $0.63$ [48]. We found that in 90.6% of the cases where a post includes an image, at least one of the replies and follow-up posts also includes an image. Further comparing the image inclusion between the three different issue types in Bugzilla, we found that **issue reports categorized as defects or enhancements had a steadily increasing trend over time in terms of the likelihood of including images,** as can be seen in Figure 5. Table 4 shows the absolute number of issue reports with and without images. Defects include the most considerable proportion of bug reports (85.8% of all the issue reports) from 2013 to 2022. Over the nine years, 46.9% of the *defects*, 22.2% of the *enhancements*, and only 1.7% of the *tasks* included an image.

The majority of the issues with an *unconfirmed* status and approximately one-third of the issues with *new* status include images, yet these differences are not statistically significant. As for the status of the issue reports, the majority (69.8%) of the issue reports are (newly) *assigned*, followed by 14.1% of *reopened* issue reports, 8.27% *unconfirmed*, and 7.81% *assigned*. Figure 6 shows the proportion of issue reports having an image within each category. *Unconfirmed* issue types had the largest proportion of issue reports including an image (53.4% of all the issues). Our initial look at a small sample of the unconfirmed issues shows the developers' effort to explain and demonstrate the problem using screenshots for bug reproduction to get the issue report confirmed. After that, the issue type with the *new* status has the highest proportion of issues including images (32.4% of the total issue reports). However, our analysis of group means did not show a significant difference between the number of shared images between different status groups (namely *new*, *assigned*, *unconfirmed*, and *reopened*).

Looking into the time it takes to close change requests on Bugzilla, we found that **issue reports including an image are closed twice as fast as other issues.** We defined $Time_{Closure}$ as the time between the initial post and the time a

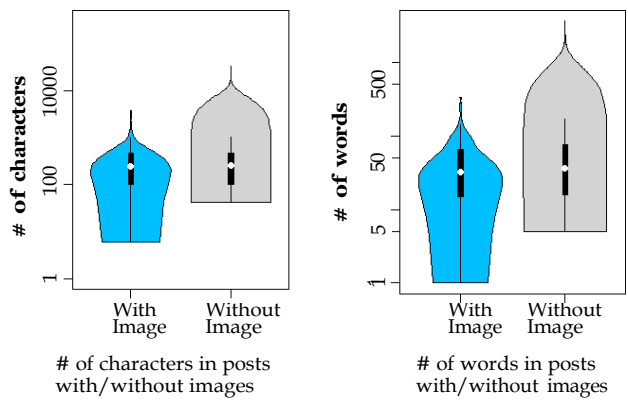

# of characters in posts
with/without images

# of words in posts
with/without images

**FIGURE 4**  The violin chart distribution of the number of characters and words in the issue reports with and without images in Bugzilla.

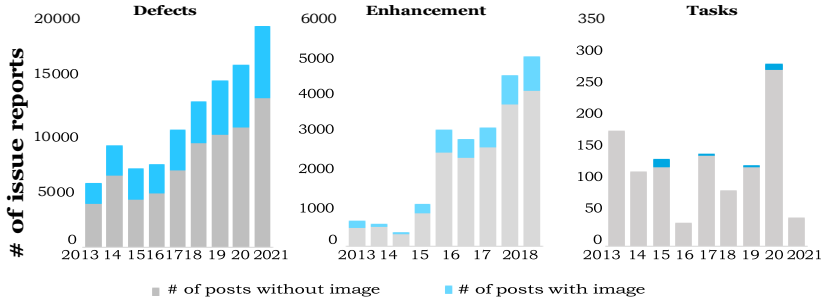

**FIGURE 5** Number of issue reports for each type (defect, enhancement, and task) with and without images.

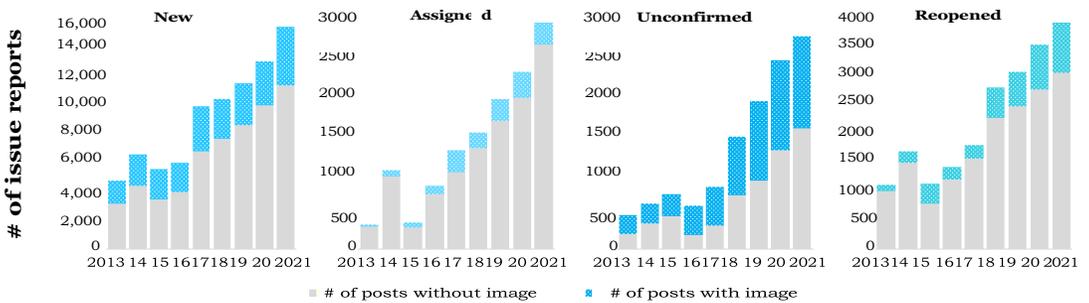

**FIGURE 6** Proportion of issue reports for each status (new, assigned, unconfirmed, reopened) with and without images.

change request is closed (Bugzilla). Then, we used the Mann-Whitney test to statistically compare the group means of $Time_{Closure}$ for the posts including an image versus the ones without images. Looking into $Time_{Closure}$ showed a significant difference in the group means in Bugzilla (p-value = 0.03, effect size = 0.33, for an $\alpha = 0.05$). In addition, On average, issue reports including an image are closed almost twice faster than those without an image.

We further compared the time, the $Time_{Closure}$, and the severity of the issues. As expected, the severe issues (i.e., severities S1 and S2) have been fixed significantly faster (p-value = 0.001, effect size = 0.15, for an $\alpha = 0.05$). We further focused on severe issues of type S1 and S2 and compared the $Time_{Closure}$ for the ones including images and those not including images. The results of a Mann-Whitney test, however, did not show any significant difference in $Time_{Closure}$ for severe issues with and without images (p-value = 0.08, Effect size = 0.15, for an $\alpha = 0.05$).

### 5.4.2 | Responsiveness on Stack Overflow

For Stack Overflow, we investigated the trend and the relation between the number of images in questions and answers between 2013 and 2022. **We observed that posts with an image in the question have significantly more images in their answer.**

Figure 7 shows the proportion and the trend for questions and for answers in Stack Overflow with and without images.

We found a strong positive (= $0.89$) Spearman correlation between the number of imaged questions and responses during these years. This observation implies a direct relationship between the number of shared images in questions and responses. We also compared the number of responses for the questions including an image and those without an image. Using a Mann-Whitney test ($\alpha = 0.05$), **we found significantly more responses to questions with an image** (p-value = 0.03).

The average length of responses for the questions sharing an image was 77.9 words compared to 93.3 words for the responses to the questions without images. When a question included an image, only 6.6% of the responses were shorter than 20 words. Similar to the Bugzilla case, here again, we looked into the length of the text for questions and answers in terms of the number of words and characters and compare those for posts with and without images.

**Posts without an image are on average 75.4% lengthier than those with an image, but the difference is not statistically significant.**

For this purpose, we counted and included the code snippets, if any, as part of the text. For the questions or answers with images, the average length of questions in our data set was 72.1 words and 362.2 characters (without space), while the average length for answers was 94.9 words and 459.3 characters. This compares to an average of 103.7 words and 604.1 characters for questions without images, and 161.9 words and 835.6 characters for answers in posts without images. Figure 8 shows the length distribution in terms of the number of words and characters for questions and answers together. The Mann-Whitney statistical test results showed no significant difference in the textual length of questions or answers with and without any image (for an $\alpha = 0.05$). We further elaborated on the relation between the accepted answers and the inclusion of images in the responses.

**Questions with an image have a significantly higher probability of accepted answers in comparison to the ones without an image.**

We evaluated if and to what extent the inclusion of the images correlates with the acceptance of the answers between the years 2013 and 2022. The results showed that 81% of the questions that include an image had accepted answers that also included an image. Furthermore, we found that there is a positive correlation of 0.73 between including an image in the answers and that answer being accepted in the Stack Overflow. We used the Mann-Whitney test to compare the number of answers for the posts with and without images. The results of our analysis showed that there is a statistically significantly higher number of accepted answers for questions including an image (p-value = 0.03 and Cliff's Delta effect size = 0.68).

We also compared the $Time_{Answer}$ and $Time_{AcceptedAnswer}$ for questions with and without images, leading to

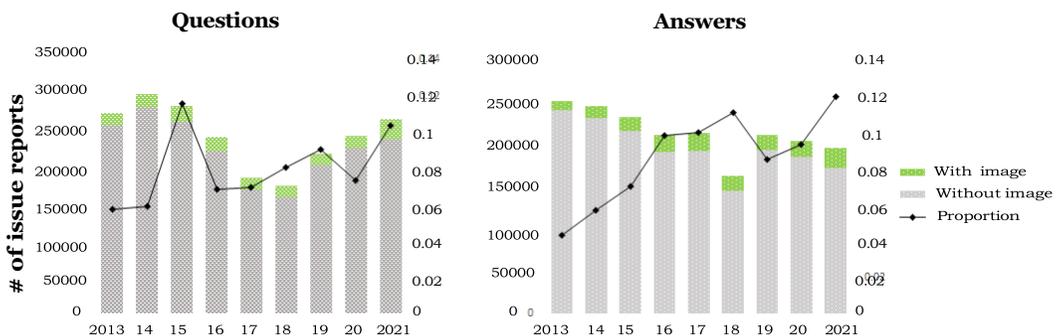

**FIGURE 7**  The trend of image sharing in questions and answers, separately, in Stack Overflow.

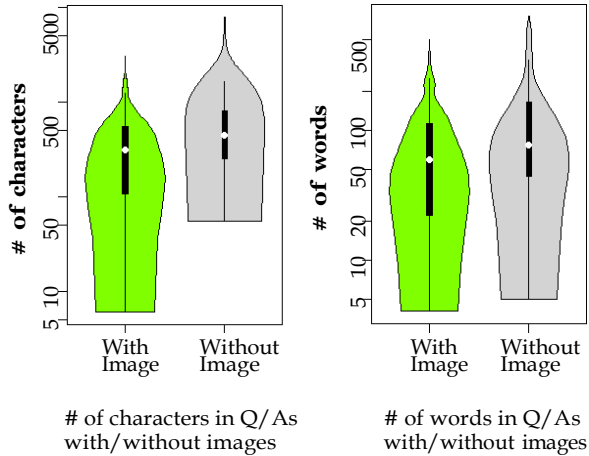

# of characters in Q/As
with/without images

# of words in Q/As
with/without images

**FIGURE 8** Number of characters and number of words in the Stack Overflow Q/A for posts with and without an image.

the finding that **questions with an image receive an accepted answer significantly (33.1%) faster than the questions without an image.**

We used the Mann-Whitney test to compare the group means statistically for the questions including images versus those without. Looking into the $Time_{AcceptedAnswer}$ with a Mann-Whitney test showed a significant difference in the group means with a p-value = $0.01$ (for an $\alpha = 0.05$) and Cliff's Delta effect size = $0.24$ (slightly larger than a small effect size). On average, questions that included an image had a 33.1% shorter $Time_{AcceptedAnswer}$. Similarly, questions with images are significantly getting their first answer faster. When comparing questions with and without images in Stack Overflow, the Mann-Whitney test for $Time_{Answer}$ returned a p-value = $0.03$ (for an $\alpha = 0.05$) and Cliff's Delta effect size = $0.33$

**FINDING 3: Image sharing correlates with stronger community engagement.** When we asked software developers in **RQ3** if they have any preference toward responding to posts with/without images, the majority (58.9%) preferred posts with images. In both Bugzilla and Stack Overflow, the number of replies and follow-up posts is significantly higher for the posts with an image (**RQ4**). In Bugzilla, the number of replies for the issue reports that include an image is, on average, two times higher than the issue reports without an image. Also, 81% of the questions that included an image in Stack Overflow also had an accepted answer. Furthermore, our observations showed a strong positive correlation between the inclusion of an image in an answer and that answer being accepted on Stack Overflow. The significant difference in the mean value of the number of replies for questions with and without an image also strengthens our conclusion that including an image correlates with higher engagement within Stack Overflow. We further observed that there is a direct relation between the number of shared images in questions and responses.

We also compared the distribution of the length of the posts with and without images in the Mozilla issue tracking system and Stack Overflow. We did not find a significant difference in the number of words or the number of characters for any of the platforms. However, in both cases, the text's average length was almost half as long when accompanied by an image. In Stack Overflow, the mean difference and the distribution are not significantly different. However, in Bugzilla, issue reports without an image are, on average 46% longer in terms of characters and 60% longer in terms of words compared to the reports with images. This implies a need for further elaboration, as our observations are not conclusive.

This shorter text description could be a confounding factor that is preventing developers from understanding the text alone without an image (Finding 2). Our analysis showed that a higher proportion of images could be found in the issue reports categorized as a *defect*, with 46.9% of the reported defects including at least one image between 2013 and 2022.

---

**So What?** The prevalence of image sharing among software developers, coupled with the strong positive correlation between images and community engagement, highlights the potential benefits of incorporating visual elements in software development processes. The majority of developers in our survey expressed a preference for posts with images, indicating that visual content can assist in conveying information more effectively. The higher number of replies and follow-up posts to issues and questions with images on platforms like Bugzilla and Stack Overflow demonstrates the value of visual aids in driving active participation and collaboration within software development communities. Considering the survey with the developers (**RQ2**) and our observation in **RQ4** we believe that there is the possibility of utilizing images as a means to enhance the clarity and understanding of complex software issues. These findings call for further exploration and analysis to fully understand the benefits and potential challenges of integrating visual elements into software development practices.

---

## 5.5 | Relation of Image Types and Other Attributes (RQ5)

We observed and reported the trend of sharing images in **RQ1** and the perception of software developers in **RQ2** and **RQ3** for 22,561 of the images. In **RQ2**, we also defined categories for the images by manually labeling 2,000 random images across Bugzilla and Stack Overflow, with sample images shown in Figure 3. While we observed trends and overall differences between issues or questions and answers with and without images, we are interested in investigating the extent to which these differences differ across image types. Hence, in this research question (**RQ5**), we analyzed the relation between the image type and seven attributes issue status, severity, text length, # of replies, type, # of accepted answers, Time_Closure), and time to get the first response upon their availability in Bugzilla or in Stack Overflow. We tested the eight hypotheses as listed in Section 3. We tested the $H_{01}$ to $H_{08}$ by Kruskal-Wallis test (see Section 3). Table 5 provides each hypothesis's p-value and effect size.

**We could not find any significant difference between the group means when testing $H_{02}$, $H_{03}$, $H_{04}$, $H_{05}$, $H_{07}$, $H_{08}$.** As a result, we could not reject any of the listed hypotheses. In other words, we did not observe any statistically significant result that indicates a relation between the nature (type) of the image and the type, status, severity, $Time_{Closure}$, # of accepted answers, or time to get a first answer. To some extent, this confirms that the different types of images are included consistently across issue types, status and severity. On the other hand, no image type stands out either in terms of shorter or longer $Time_{Closure}$, which corroborates the lack of significant differences in terms of issue severity. Similarly, while more answers are found for Stack Overflow posts with images, the lack of significance in terms of the number of accepted answers or the time to get a first answer prevents us from any reasoning here.

**TABLE 5** Results of the Kruskal–Wallis test for the tested hypothesis in **RQ5**.

| Null hypothesis | P-Value | Effect size | Decision |
|---|---|---|---|
| $H_{01}$ | 0.021 | 0.71 | Rejected |
| $H_{02}$ | 0.191 | 0.78 | Not rejected |
| $H_{03}$ | 0.263 | 0.66 | Not rejected |
| $H_{04}$ | 0.076 | 0.80 | Not rejected |
| $H_{05}$ | 0.064 | 0.79 | Not rejected |
| $H_{06}$ | 0.018 | 0.82 | Rejected |
| $H_{07}$ | 0.209 | 0.58 | Not rejected |
| $H_{08}$ | 0.780 | 0.83 | Not Rejected |

However, we found a significant difference in the group means for the *# of replies in Bugzilla* ($H_{01}$) and *# of answers in Stack Overflow* ($H_{06}$). A rejected null hypothesis of the Kruskal-Wallis test indicates that at least the mean of one group (here, we have ten groups each for an image category) is different from the others. As a result, we need to follow post-hoc tests to identify between the ten image types which group(s) has a significantly different # of replies in $H_{01}$ and median of the # of answers in $H_{06}$. We used Dunn post-hoc tests [48] for this. According to this test, the group size can be equal or unequal, and at each level, the p-value is adjusted for multiple comparisons.

**We found a significant difference in the # of replies on Bugzilla posts including a "Run time error" image.** As a result, we find that the # of replies in $H_{01}$ for "Run time error" is significantly higher than all the other groups with the average adjusted p-value = $0.004$. Also, **we observed a significant difference in the # of answers in Stack Overflow containing a "Run time error" or "Dialog box" image.** When using Dunn's test for $H_{06}$, our findings show a significantly higher mean for the # of answers in Stack Overflow containing "Run time error" or "Dialog box" images compared to all other groups (average p-value of $< 001$ and $= 0.005$ respectively). We did not find any significant difference between these two groups (p-value = $0.058$) on Stack Overflow when testing $H_{06}$. This implies that the group means of # of answers between "Dialog box" and "Run time error" posts in our sample are not significantly different.

<u>**FINDING 4:**</u> **Sharing images correlates with reduced response time and improved developer communication -** In both platforms, issues including an image are being responded to at least 30% faster than the others. For the issue reports of the studied Mozilla projects, the ones including an image are closed almost twice faster than the reports without an image, and we found a significant difference between the closure time for tickets with and without an image. In Stack Overflow, we also found significantly lower response time for the questions that include an image. This involved a faster first response and a shorter time to get an answer accepted. The time to get the first response is shorter for 33.2% of the questions with an image when compared to questions without any image.

Images correlate with a faster reaction by developers and a better understanding. Our survey showed that 89.9% of the developers share images along with their posts to facilitate comprehension. Getting an answer faster was the second most popular reason among the surveyed developers (see Table 3). We compared the time between posting a question or change request and the time it is accepted or closed for posts with versus without images (**RQ4**). We found that in both Stack Overflow and Bugzilla, the time is significantly shorter for posts that include images. Sharing images is trending and seems to help close questions/bug reports faster and facilitate developers' understanding.

---

**So What?** *Software engineering crowd-sourcing platforms should consider integrating tools and techniques for facilitating image sharing and processing.* Software development is a communication-intensive process that requires understanding, synchronization, and discussion of information. So far, the evolution of tools and techniques for assisting software developers has been dominated by understanding and generating only the textual content [6, 7, 8, 9]. The growing number of images developers share provides a unique opportunity to train machines on reusing and even synthesizing new images to make the software artifacts visual. In the future, automation tools should be extended to process posted images along with natural and programming language processing. The increasing trend in sharing images and developers' engagement allows crowd-sourcing and social coding platforms to engage users and potentially improve communication and reduce turnaround times by supporting image sharing and automated processing.

# 6 | LIMITATIONS AND THREATS TO VALIDITY

As the first study on analyzing the use of images by developers in social coding platforms, the paper stays descriptive. We foresee future research that focuses on developers' productivity [52, 53], including diary analysis that would further delve into the habitual aspects [54] of sharing images and their relation with developers' behavioral and professional attributes. The current tools for automated mining of developers' behavior and improving productivity in software teams are based on mining activities or textual artifacts. However, in consideration of the increasing trend of sharing images, we foresee a change in the development of automated tools. We discuss threats to the validity of our observations and the limitations of this study following the categorization provided by Wohlin et al. [45].

**Conclusion validity-***Are we drawing the correct conclusion about treatment and outcome relation?*: The nature of this study is exploratory and observational as it is one of the first studies in the domain. We primarily rely on descriptive statistics to discuss the status quo of image sharing in social coding environments. While we discussed and used the standard p-value and side effect values, one should be cautious about not deriving causality from these observations.

**Construct validity -***Are we measuring the right things?* The data for the first four research questions have been gathered over nine years and across two platforms. While this might not represent all the platforms hosting software activities, we triangulated our findings with the direct survey with developers to reduce bias in our results. We could gather responses from 167 developers, which is considered a good size for a survey considering the state of the art in software engineering. Of course, as we are discussing third-degree data in the first four RQs, they might have been impacted by confounding factors that have not been considered in our study. We followed the guidelines of Kitchenham and Pfleeger [46] for constructing the surveys and further followed a democratic approach where more than one annotator should answer each question. Nevertheless, there might be some bias toward a positive answer or a random response in the survey we have performed with the developers.

**Internal Validity -***Can we be sure that the treatment indeed caused the outcome?* As we had two controlled groups of questions/issues with clear selection criteria (with an image vs. without an image), the risk of violating internal validity appears to be low.

**External Validity -***Can the results be generalized beyond the scope of this study?* We ran two parallel case studies on the existence and trends of image sharing in Bugzilla and in Stack Overflow. Bugzilla has long been the default repository for research on defect prediction and resolution as well as the analysis of developers' productivity and collaboration in open source. Stack Overflow is also a well-known platform among developers to crowdsource answers to their technical issues, and the research community has been utilizing and mining this repository to increase productivity in technical teams.

In addition, we believe that the popularity and the extended nine-year analysis of image data represent the existing trends. The majority of our observations were similar across both Bugzilla and Stack Overflow. Furthermore, the developers' survey confirmed our findings in terms of the value and complementary nature of the shared images. All in all, our findings are aligned with the trends in general-purpose social media platforms. Considering that a considerable portion of studies on social coding platforms has been conducted on Stack Overflow and Bugzilla, we believe this trend and analysis are fairly generalizable. Future work should consider other platforms to strengthen the generalizability of the results.

# 7 | CONCLUSION

Instagram and Pinterest are prominent examples of visual social networks. While limiting the size of the textual content, their primary intent is to share images. Considering the social nature of software development and its impact on development practices, we can expect that developers would lean toward repeating the behavior met in other social media, exploiting the power of visual communication where possible. For example, Stack Overflow and Bugzilla allow users to attach additional files along with their posted content. Taking these two channels as a widely acknowledged medium for software maintenance and knowledge management activities, we examined the trend of sharing visual content in these two channels. We further triangulated our observations by surveying 168 software developers by actively evaluating 22,561 images. The increasing trend of sharing images and our observed benefits for getting faster and more precise responses led us to predict that the future of the software development team would require advances in research and practice to automate the analysis of the visual data shared by developers. As a result, the tools and techniques developed and commonly present to increase developers' productivity would further need to be extended to use Natural Language Processing and involve Image Processing techniques.

In the past, we as a community have been able to successfully summarize software artifacts (bug reports, source code, mailing lists, developer discussions) into shorter text [55]. The community went even further and automatically generated textual commit messages, release notes, pull requests, or replies to Stack Overflow questions, and built chatbots by relying on analogies and natural language processing techniques [56]. In the future, the developers' knowledge-sharing and communication assistance tools (such as the ones named above) should be expected to complement the current text-based artifacts by synthesizing and generating images. These images will be either retrieved and reused from shared images or will be synthesized and newly generated by automated techniques. This future is accessible and mainly encouraged by the success in teaching machines sophisticated features such as humans' facial and body attributes, or attributes of transportation means and driveways for self-driving cars. Hence, it is quite probable that automated tools for helping software developers achieve high precision on generated images. Software interfaces, inputs, and outputs are less geometrically complex and with less variety in comparison to solved problems in the domain of image processing in self-driving cars. To conclude, there is a clear opportunity for machines to learn and synthesize software-related images to support software development tasks.